\documentclass[doublecol]{epl2} 
\usepackage{graphicx}
\usepackage{mathrsfs}
\usepackage[T1]{fontenc} 
\usepackage{bbold}
\usepackage{url}
\usepackage{hyperref}
\usepackage[english]{babel}
\usepackage[utf8]{inputenc}
\usepackage{amsmath}
\usepackage{color}
\usepackage{amsfonts}
\usepackage{amssymb}
\usepackage{mathtools}
\usepackage{placeins}
\usepackage{float}
\usepackage{tabularx}
\usepackage{adjustbox}
\usepackage{caption}
\usepackage{subcaption}
\title{\boldmath Inflationary $\alpha$-Attractor From Type IIB/F Theory}
\shorttitle{\boldmath  Inflationary $\alpha$-Attractor From Type IIB/F Theory} 
\author{Arunoday Sarkar\inst{} \and Buddhadeb Ghosh \inst{}}
\institute{                    
  \inst{} Centre of Advanced Studies, Department of Physics, The University of Burdwan,\\Burdwan 713 104, India
  
}

\abstract{We derive an $\alpha=1/3$ -- attractor potential of slow-roll inflation in the geometric set-up of three intersecting $D7$ branes under $T^6/Z_N$ type of $CY_3$-compactification within type IIB/F-theory with some near-conifold regions. The underlying quadratic structure of the kinetic poles is found to arise from a correction in the K\"{a}hler potential when an extra contribution of open string moduli is turned on. While the closed string sector of the moduli spectrum is completely stabilized via quantum corrections of perturbative and non-perturbative origin, the open string sector plays the lead role in driving the inflationary expansion in the radial direction. A generic asymptotic behavior of the inflaton field near the pole boundaries manifests as the slow-roll plateau in canonical field space, which becomes responsible for giving universal predictions of the cosmological parameters. We find that the presence of the open strings near conifold regions brings the realization of \textit{pole inflation} in the present set up. Finally we compare our results with similar models and discuss the importance of exploring precise values of $\alpha$ in the light of ongoing and forthcoming cosmological surveys.}

\begin{document}

\maketitle
\flushbottom
Precision studies of cosmic microwave background (CMB) anisotropies and polarizations reveal that our universe is homogeneous at large scale and inhomogeneous at small scale, roughly within $100$ Mpc \cite{Dodelson:2003ft}. The large-scale structures result from gravitational amplifications of cosmological perturbations in the early universe, originating from quantum fluctuations during the period of inflationary expansion. According to Planck-2018 \cite{Planck:2018jri}, the primordial scalar perturbations are nearly scale-invariant, adiabatic, and almost Gaussian, associated with a single field slow-roll inflaton potential. The slow-roll plateau needs to have two essential properties. First, it should protect the underlying effective theory from the issues concerning infinite or superplanckian field excursion \cite{Dimopoulos:2022wzo,Dimopoulos:2017zvq} of the inflaton field. Second, it should be a generic one \textit{i.e.} independent of the fine-tuning of the model parameters such that it will ultimately be responsible for the universal predictions of the cosmological parameters \cite{Planck:2018vyg}. Apart from these features, the slow-roll plateau has some unresolved issues with the swampland distance conjecture and weak gravity conjecture. (See Ref. \cite{Scalisi:2018eaz} for detailed discussions on these topics.)\par Interestingly, a certain class of inflationary models is efficacious in conforming to all the requirements mentioned above. These potentials are the so called \textit{$\alpha$-attractors} \cite{Kallosh:2013yoa,Roest:2015qya}, best known for their universal predictions of the scalar spectral index $n_s=1-\frac{2}{N}$ and the tensor-to-scalar ratio $r=\frac{12\alpha}{N^2}$ for a large number of remaining e-folds $N$. The underlying reason behind this attractor nature is the presence of a quadratic pole in the Laurent series expansion of the coefficient of the kinetic term in the Lagrangian \cite{Galante:2014ifa}. In fact, the parameter $\alpha$ is connected with several compelling features \textit{viz.,} the double pole behavior \cite{Kallosh:2014rga,Galante:2014ifa}, asymptotic freedom \cite{Kallosh:2016gqp}, hyperbolic half-plane disc geometry \cite{Kallosh:2015zsa,Carrasco:2015uma,Carrasco:2015rva}, $B$ modes \cite{Canas-Herrera:2021sjs,Kallosh:2019hzo,Kallosh:2019eeu}, dark energy \cite{Dimopoulos:2017zvq} and the early dark energy \cite{Brissenden:2023yko,Dimopoulos:2023tcc}. Refs. \cite{Sarkar:2021ird,Sarkar:2023cpd,Sarkar:2023vpn} contain more references on these topics.\par The origin of the $\alpha$ attractor is rooted in $\mathcal{N}=1$, $d=4$ supergravity theories \cite{Kallosh:2013yoa,Roest:2015qya}. In these frameworks the inflaton field is identified as a part of a chiral super-multiplet and the pole determining $\alpha$ factor is related to the inverse curvature of the $SU(1,1)/U(1)$ K\"{a}hler manifold. To find a more fundamental origin of the inflationary $\alpha$-attractor, it should be prescribed under a theory, which is a low-energy effective description of some UV-complete theory of quantum gravity, such as string theory.\par Starting from KKLT \cite{Kachru:2003aw}, connecting the paradigm of string theory compactification with the inflationary picture has proved to be a highly non-trivial task. In a number of contemporary literature, such as \cite{Antoniadis:2020stf,Basiouris:2020jgp,Basiouris:2021sdf,Let:2022fmu,Let:2023dtb}, this effort has resulted in finding an effective potential in four dimension through K\"{a}hler moduli stabilizations by perturbative as well as non-perturbative quantum corrections. This low-energy effective potential serves as the (slow-roll) inflaton potential in cosmology. From string theoretic perspective, this effective potential is connected with the moduli stabilization problem. However, from the cosmological point of view, it fails to have a generic slow-roll plateau, in order to address the universal predictions of cosmological parameters which are very essential for obtaining an experimentally viable inflaton potential, as mentioned earlier. This is because, in all such prescriptions the effective Lagrangians lack the presence of the kinetic poles. Consequently, the resulting potentials do not show the attractor behaviour. Therefore, in order to derive an inflaton potential of attractor type like the $\alpha$-attractor, the Lagrangian should be equipped with the poles in its kinetic term.\par In the present paper, our aim is to find the pole structure in an effective Lagrangian $-$ precisely the quadratic poles of the $\alpha$-attractor $-$ derived from the string theory compactifications. This could be a novel means of connecting the stringy perspectives with the inflationary scenario. We develop our formulation by modifying the moduli stabilization mechanism discussed in Refs. \cite{Let:2022fmu,Let:2023dtb} with some new elements from the open string sector and the internal geometry of the underlying manifold structure, which will bring the required poles in the Lagrangian.\par We start by considering the type IIB/F-theory, compactified on the $T^6/Z_N$ orbifold limit of $CY_3$ in the geometric set-up of three intersecting $D7$ branes equipped with $NS$ and $RR$ closed three form world volume fluxes from $D_{1,3,5,7}$ branes. The \textit{pole inflation} is found to be not accessible from closed string moduli and large volume scenario alone \cite{Dias:2018pgj} as in Refs. \cite{Let:2022fmu,Let:2023dtb}. Therefore, in addition to the axion-dilaton ($S$), complex structure ($z$) and the K\"{a}hler ($\rho$) moduli in the closed string sector, here, we invoke some special structures \textit{viz.,} the near-conifold-type singular regions and some $N$ moduli $\Phi_i\in \mathbb{C}$ defined on a unit disk and associated with the open strings attached to the $D3$ branes present in the near-conifold regions. The relevant K\"{a}hler potential takes the form,
\begin{equation}\small
\begin{split}\small
    \mathcal{K}^{(0)}=&\sum_{j=1}^2-\ln{\left[-i\left(\rho_j-\Bar{\rho}_j\right)\right]}-\ln{\left[-i\left(\rho_3-\Bar{\rho}_3\right)-\sum_{k=1}^N\Phi_k\Bar{\Phi}_k\right]} \\
    &-\ln{\left[-i\left(S-\Bar{S}\right)\right]}-\ln{\left(-i\int\Omega(z)\wedge \Bar{\Omega}(\Bar{z})\right)},
\end{split} 
\label{eq:1}
\end{equation} where $\Omega(z)$ is a holomorphic $(3,0)$ form. In terms of the $CY_3$ volume $\mathcal{V}=(\Pi_{i=1}^3\tau_i)^{1/2}$, $\tau_i$ being the $4$-cycle volume\footnote{Actually, $\rho_k = b_k+i\tau_k$, where $b_k$ are axion moduli. In the present formalism, we put $b_k=0$, for simplicity.}, $\mathcal{K}^{(0)}$ can be expressed as,
\begin{equation}\small
\begin{split}\small
    \mathcal{K}^{(0)}=&-2\ln{\mathcal{V}}-\ln{\left(1-\frac{\sum_{k=1}^N \Phi_k\Bar{\Phi}_k}{2\tau_3}\right)}\\&-\ln{\left[-8i\left(S-\Bar{S}\right)\right]}-\ln{\left(-i\int\Omega(z)\wedge \Bar{\Omega}(\Bar{z})\right)}.
\end{split}
\label{eq:volumeKahler}
\end{equation}
We have assumed that the near-conifold region is formed by the warping of largest $4$-cycle, belonging to $\tau_3$. Thus, the  open string moduli $\Phi_k$ contribute to $\tau_3$ only. This tree-level K\"{a}hler potential together with a tree-level superpotential $\mathcal{W}_0$ provides a low-energy effective description of type  IIB string theory, called $\mathcal{N}=1$ type IIB supergravity. The corresponding $SL(2,\mathbb{R})$ invariant Lagrangian \cite{Giddings:2001yu} contains an $F$-term potential \cite{Freedman:2012zz}
\begin{equation}\small\small
V_F=e^{\mathcal{K}^{(0)}}\left({\mathcal{K}^{(0)}}^{\alpha\Bar{\beta}}D_{\alpha}\mathcal{W}_0 D_{\Bar{\beta}}\Bar{\mathcal{W}_0}-3\mathcal{W}_0\Bar{\mathcal{W}_0}\right),
\end{equation}
as an effective potential and a kinetic part 
\begin{equation}\small\small
\mathcal{L}_{\mathrm{kin}}=\mathcal{K}^{(0)}_{\alpha\Bar{\beta}}\partial Z^{\alpha}\partial\Bar{Z}^{\Bar{\beta}},
\label{eq:kkin}
\end{equation}
which is determined by the K\"{a}hler metric components $\mathcal{K}^{(0)}_{\alpha\Bar{\beta}}=\partial_{\alpha}\partial_{\Bar{\beta}}\mathcal{K}^{(0)}$ and the $Z^{\alpha}$ coordinates of the $CY_3$ moduli space. Using Eq. (\ref{eq:1}) we obtain the $\mathcal{K}^{(0)}_{\alpha\Bar{\beta}}$ elements for $\rho$, $S$ and $\Phi_k$ as,
\begin{equation}\small
\begin{split}\small
    &\mathcal{K}^{(0)}_{\rho_1\Bar{\rho}_1}=-\frac{1}{(\rho_1-\Bar{\rho}_1)^2},\quad\quad\mathcal{K}^{(0)}_{\rho_2\Bar{\rho}_2}=-\frac{1}{(\rho_2-\Bar{\rho}_2)^2},\\
    &\mathcal{K}^{(0)}_{\rho_3\Bar{\rho}_3}=\left[-i\left(\rho_3-\Bar{\rho}_3\right)-\sum_{k=1}^N \Phi_k\Bar{\Phi}_k\right]^{-2},\mathcal{K}^{(0)}_{S\Bar{S}}=-\frac{1}{(S-\Bar{S})^2},\\
    &\mathcal{K}^{(0)}_{\rho_3\Bar{\Phi}_j}=i\Phi_j\left[-i\left(\rho_3-\Bar{\rho}_3\right)-\sum_{k=1}^N \Phi_k\Bar{\Phi}_k\right]^{-2},\\
    &\mathcal{K}^{(0)}_{\Phi_j\Bar{\rho}_3}=-i\Bar{\Phi}_j\left[-i\left(\rho_3-\Bar{\rho}_3\right)-\sum_{k=1}^N \Phi_k\Bar{\Phi}_k\right]^{-2},\\
    &\mathcal{K}^{(0)}_{\Phi_j\Bar{\Phi}_{j'}}=\frac{\left(\Bar{\Phi}_j\Phi_{j'}+\delta_{jj'}\left(-i\left(\rho_3-\Bar{\rho}_3\right)-\sum_{k=1}^N \Phi_k\Bar{\Phi}_k\right)\right)}{\left[-i\left(\rho_3-\Bar{\rho}_3\right)-\sum_{k=1}^N \Phi_k\Bar{\Phi}_k\right]^{2}},
\end{split}
\label{eq:kcomponents}
\end{equation} and that for $z$ is determined by the line element of complex structure moduli space, given in Ref. \cite{Giddings:2001yu}. In spherical polar coordinates $(r, \theta, \sigma)$, the $\Phi_k$ moduli can be expressed in terms of the radial and the angular components as, 
\begin{equation}\small\small
    \Phi_k=\mathcal{R}(r)\Theta_k(\theta) e^{-i\gamma_k(\sigma)},
    \label{eq:phimoduli}
\end{equation}
such that $\sum_{k=1}^N \Theta_k^2 =1$ \cite{Dias:2018pgj} and therefore $\sum_{k=1}^N |\Phi_k|^2 = \mathcal{R}^2$. Here, we assume that all $D3$ branes are arranged in such a way that, the variations of $\Phi_k$ moduli along the radial direction behave as fields, which are identical for all the $D3$ branes, and also we do not consider any variations along the angular directions. This we do, in order to maintain a single-field picture. Therefore, $\Theta_k$ and $\gamma_k$ are fixed for each $D3$ brane, confined in a conifold singularity. $\Theta_k$'s are treated as some basis functions, whose properties have been described in Ref. \cite{Dias:2018pgj}. \par From Eqs. (\ref{eq:kkin}), (\ref{eq:kcomponents}), (\ref{eq:phimoduli}) and the two equations following Eq. (\ref{eq:phimoduli}), the kinetic part of Eq. (\ref{eq:kkin}) becomes,
\begin{equation}\small
   \begin{split}\small
\mathcal{L}_{\mathrm{kin}}=&\sum_{i=1}^2 \frac{(\partial\tau_i)^2}{4\tau_i^2}+\frac{(\partial\tau_3)^2-(\partial\tau_3)(\partial \mathcal{R}^2)+2\tau_3(\partial\mathcal{R})^2}{(2\tau_3-\mathcal{R}^2)^2}\\
&-\frac{\partial S\partial\Bar{S}}{(S-\Bar{S})^2}+\sum_{\mu,\Bar{\nu}=1}^{h^{2,1}}\mathcal{K}^{(0)}_{\mu\Bar{\nu}}\partial z^{\mu}\partial\Bar{z}^{\Bar{\nu}}.
   \end{split}
   \label{eq:3}
\end{equation} As is well-known, $S$ and $z$ are stabilized through tree-level superpotential $\mathcal{W}_0$ at the classical level \cite{Giddings:2001yu}. Thus the third and the fourth terms in Eq. (\ref{eq:3}) are constant classical backgrounds and we can, effectively, write,
\begin{equation}\small
\mathcal{L}_{\mathrm{kin}}=\sum_{i=1}^2 \frac{(\partial\tau_i)^2}{4\tau_i^2}+\frac{(\partial\tau_3)^2-(\partial\tau_3)(\partial \mathcal{R}^2)+2\tau_3(\partial\mathcal{R})^2}{(2\tau_3-\mathcal{R}^2)^2},
   \label{eq:3modified}
\end{equation} because, the derivatives of $S$ and $z$ have vanished due to stabilization. Now, we allow two KKLT-type non-perturbative corrections in $\mathcal{W}_0$ through $\tau_{1,2}$ corresponding to smaller $4$-cycles as
\begin{equation}\small
 \mathcal{W}=\mathcal{W}_0 + \sum_{i=1}^2 B_i\exp{\left(-b_i\tau_i\right)}  
\end{equation}
in addition to the perturbative volume correction up to one-loop order in the K\"{a}hler moduli dependent part of $\mathcal{K}^{(0)}$ in Eq. (\ref{eq:volumeKahler}) as
\begin{equation}\small
   \mathcal{K} (\mathcal{V},\Phi_k)= -2\ln{\left(\mathcal{V}+\xi+2\eta\ln{\mathcal{V}}\right)-\ln{\left(1-\frac{\sum_{k=1}^N \Phi_k\Bar{\Phi}_k}{2\tau_3}\right)}}.
\end{equation} The terms containing $\xi$ and $\eta$ indicate the tree-level $\alpha'$ correction \cite{Becker:2002nn} and the one-loop correction \cite{Antoniadis:2018hqy,Antoniadis:2019rkh} respectively. Then imitating a process as in Ref. \cite{Let:2023dtb}, $\tau_{1,2}$ can be stabilized into constant values, say $\langle\tau_{1,2}\rangle$, making the volume $\mathcal{V}$ effectively dependent on $\tau_3$. The resulting non-vanishing effective $dS$ potential
\begin{equation}\small
V(\mathcal{V},\Phi_k)\equiv V(\mathcal{V},\mathcal{R})= V_{F}(\mathcal{V},\mathcal{R})+V_D(\mathcal{V})
\label{eq:6}
\end{equation}
is the sum of $F$-term potential
\begin{equation}\small\small
    V_F(\mathcal{V},\mathcal{R})=e^{\mathcal{K}}\left(\mathcal{K}^{\alpha\Bar{\beta}}D_{\alpha}\mathcal{W}D_{\Bar{\beta}}\Bar{\mathcal{W}}-3\mathcal{W}\Bar{\mathcal{W}}\right),
\end{equation}
 where $D_{\alpha}\equiv\partial_\alpha+\partial_\alpha\mathcal{K}$ is the connection over $CY_3$ moduli space and the $D$-term potential
\begin{equation}\small\small
     V_D(\mathcal{V})=\sum_{i=1}^2\frac{d_i}{\langle\tau_i\rangle^3}+d_3\frac{(\langle\tau_1\rangle\langle\tau_2\rangle)^3}{\mathcal{V}^6}.
\end{equation}
Eq. (\ref{eq:6}) shows that the effective potential is a two-field potential, but, as required, we need a single field attractor potential, which can be obtained by stabilizing the volume at $\mathcal{V}=\langle\mathcal{V}\rangle$, where the potential attains a minimum in $\mathcal{V}$ direction \textit{i.e.} ,
\begin{equation}\small
    \frac{\partial V(\mathcal{V},\mathcal{R})}{\partial\mathcal{V}}|_{\mathcal{V}=\langle\mathcal{V}\rangle}=0\quad\quad\mathrm{and}\quad\quad\frac{\partial^2 V(\mathcal{V},\mathcal{R})}{\partial\mathcal{V}^2}|_{\mathcal{V}=\langle\mathcal{V}\rangle}>0.
\end{equation}
Thus Eq. (\ref{eq:6}) becomes
\begin{equation}\small
    V(\mathcal{R})\equiv V(\langle\mathcal{V}\rangle,\mathcal{R})=V_F(\langle\mathcal{V}\rangle,\mathcal{R})+V_D(\langle\mathcal{V}\rangle),
\end{equation} where $V_D(\langle\mathcal{V}\rangle)$ acts as a constant uplifting factor from $AdS_4$ to $dS_4$ space. In this way, $\tau_3$ is, actually, stabilized to the value,
\begin{equation}\small\small
    \langle\tau_3\rangle=\frac{(\langle\mathcal{V}\rangle)^2}{\langle\tau_1\rangle\langle\tau_2\rangle}.
\end{equation}
Now, from Eq. (\ref{eq:3modified}) the kinetic part can be written in simple form after stabilizations of all the $\tau_i$ moduli as,
\begin{equation}\small\small
\mathcal{L}_{\mathrm{kin}}=\frac{4\langle\tau_3\rangle}{(2\langle\tau_3\rangle-\mathcal{R}^2)^2}\frac{\left(\partial\mathcal{R}\right)^2}{2}.
\label{eq:9}
\end{equation}
After stabilizations of all the closed-string moduli, $\mathcal{L}_{\mathrm{kin}}$ represents a kinetic term in the usual four dimensions, where $\mathcal{R}$ behaves like a field with a potential $V(\mathcal{R})$. The corresponding effective Lagrangian in complete form can be written including the kinetic part and the effective potential as,
\begin{equation}\small\small
    \mathcal{L}=\sqrt{-g}\left[\frac{R_4}{2}-\frac{4\langle\tau_3\rangle}{(2\langle\tau_3\rangle-\mathcal{R}^2)^2}\frac{\left(\partial\mathcal{R}\right)^2}{2}-V(\mathcal{R})\right],
    \label{eq:10}
\end{equation} where $g$ and $R_4$ are the metric and the Ricci scalar in ordinary four dimensions. Eqs. (\ref{eq:9}) and (\ref{eq:10}) highlight that a quadratic pole structure at $\mathcal{R}=\pm\sqrt{2\langle\tau_3\rangle}$ has arrived in the kinetic part, which is entirely controlled by the perturbatively stabilized volume modulus $\tau_3$. The field $\mathcal{R}$ \textit{emerges} as a non-canonical inflaton field in Einstein frame, coupled minimally with gravity\footnote{Actually, the steps by which one arrives at Einstein frame from string frame, are quite elaborate \cite{Giddings:2001yu}. However, in this article, we are concerned with the scalar inflaton potential only.} through the derivative factor. The poles restrict the inflaton field within the pole boundary, making the field displacement $\Delta\mathcal{R}$ to be sub-Planckian ($\Delta\mathcal{R}<M_p$, $M_p$ being the reduced Planck mass) and therefore the effective theory becomes safe from being plagued by radiative corrections \cite{Dimopoulos:2017zvq,Dimopoulos:2022wzo}. Generally, when the inflaton field tends to encounter the poles, quantum corrections become large and sometimes uncontrollable. But as discussed in Ref. \cite{Dias:2018pgj}, the $\Phi_k$ moduli significantly suppress the growth of the radiative effects near the boundary of the moduli space as $N$ becomes large ($\sim 10^4$). This will depend on the numbers of $D3$ branes and the open strings attached to them, which are fixed by the charge conserving tadpole condition \cite{Giddings:2001yu}. A huge population of open string moduli is, therefore, favorable for obtaining an effective theory of inflation in contrast to the limited number of closed string moduli. \par Now, expanding the term containing the poles about $\mathcal{R}=\sqrt{2\langle\tau_3\rangle}$ of Eq. (\ref{eq:10}) by Laurent series \cite{arfken2011mathematical} as,
\begin{equation}\small\small
\begin{split}\small\small
     \frac{4\langle\tau_3\rangle}{(2\langle\tau_3\rangle-\mathcal{R}^2)^2}=&\frac{1}{2(\mathcal{R}-\sqrt{2\langle\tau_3\rangle})^2}-\frac{1}{2\sqrt{2\langle\tau_3\rangle}(\mathcal{R}-\sqrt{2\langle\tau_3\rangle})}\\&+\sum_{n=0}^{\infty}a_n(\mathcal{R}-\sqrt{2\langle\tau_3\rangle})^n,
\end{split}
\label{eq:expansion}
\end{equation}
we see from the first term that the order of the leading pole is 2 and, therefore, it is a quadratic pole. The presence of the pole structure ensures that the potential $V(\mathcal{R})$ is an \textit{inflationary attractor}. The interesting feature of an attractor potential is that without going into the details of its structure or derivation, the cosmological parameters can be predicted in terms of the order of the leading pole and the coefficient (which is $\frac{1}{2}$, here) of the term containing the leading pole. For example, the inflationary spectral index $n_s$ and the tensor-to-scalar ratio $r$ can be obtained from the formulae, derived in Ref. \cite{Galante:2014ifa}, when the number of remaining e-folds $N$ is large:
\begin{equation}\small\small
 n_s=1-\frac{2}{(2-1)N}=1-\frac{2}{N} 
 \label{eq:12}
\end{equation} and 
\begin{equation}\small\small
    r=\frac{8\left(\frac{1}{2}\right)^{\frac{1}{2-1}}}{(2-1)^{\frac{2}{2-1}}N^{\frac{2}{2-1}}}=\frac{4}{N^2}.
    \label{eq:13}
\end{equation}
The total Lagrangian of Eq. (\ref{eq:10}) can be rewritten in terms of a new field, $\varphi=\sqrt{\frac{3}{2\langle\tau_3\rangle}}\mathcal{R}$ as
\begin{equation}\small\small
 \mathcal{L}=\sqrt{-g}\left[\frac{R_4}{2}-\frac{\frac{1}{3}}{\left(1-\frac{\varphi^2}{3}\right)^2}\left(\partial\varphi\right)^2-V\left(\sqrt{\frac{2\langle\tau_3\rangle}{3}}\varphi\right)\right].  
 \label{eq:14}
\end{equation}
Interestingly, the results of Eqs. (\ref{eq:12}) $-$ (\ref{eq:14}) perfectly match with that of the universal equations and the associated Lagrangian of the $\alpha$-attractor for $\alpha=\frac{1}{3}$ \cite{Kallosh:2013yoa}. Therefore we can infer that the potential in Eq. (\ref{eq:14}) is a generic $\alpha$-attractor potential with $\alpha=\frac{1}{3}$. Now, if we perform a canonical normalization by $\varphi=\sqrt{3}\tanh{\frac{\phi}{\sqrt{2}
}}$ then, we get,
\begin{equation}\small\small
   \mathcal{L}=\sqrt{-g}\left[\frac{R_4}{2}-\frac{1}{2}\left(\partial\phi\right)^2-V\left(\sqrt{2\langle\tau_3\rangle}\tanh{\frac{\phi}{\sqrt{2}}}\right)\right], 
   \label{eq:15}
\end{equation}
which can also be obtained from Eq. (\ref{eq:10}) by choosing $\mathcal{R}=\sqrt{2\langle\tau_3\rangle}\tanh{\frac{\phi}{\sqrt{2}}}$. Eq (\ref{eq:15}) can be identified as the usual Lagrangian of the canonical inflaton field $\phi$ minimally coupled with gravity. Depending upon the specific choice of dependence of $V(\mathcal{R})$ on $\mathcal{R}$,  the form of the potential will be different. For example, if a polynomial or power-law type dependence is considered \textit{i.e.}, $V(\mathcal{R})=A_{2n}\mathcal{R}^{2n}$, $n$ being a positive integer number, then,
\begin{equation}\small\small
 V(\mathcal{R}) = V\left(\sqrt{2\langle\tau_3\rangle}\tanh{\frac{\phi}{\sqrt{2}}}\right)=V_0\tanh^{2n}{\frac{\phi}{\sqrt{2}}},
\end{equation}
which is the $T$ model $\alpha=\frac{1}{3}$-attractor potential \cite{Kallosh:2015lwa}. On the other hand, if we choose $V(\mathcal{R})=A_{2n}\left(\frac{2\sqrt{2\langle\tau_3\rangle}\mathcal{R}}{\sqrt{2\langle\tau_3\rangle}+\mathcal{R}}\right)^{2n}$, then,
\begin{equation}\small\small
    V(\mathcal{R}) = V\left(\sqrt{2\langle\tau_3\rangle}\tanh{\frac{\phi}{\sqrt{2}}}\right)=V_0\left(1-e^{-\sqrt{2}\phi}\right)^{2n},
\end{equation}
which is the $E$ model $\alpha=\frac{1}{3}$-attractor potential \cite{Kallosh:2015lwa}. Here, the constant $V_0=A_{2n}(\sqrt{2\langle\tau_3\rangle})^{2n}$ sets the inflationary energy scale. In cosmology, this can be determined by the COBE/Planck normalization formula derived in Ref. \cite{Sarkar:2023cpd}, which is used to satisfy the required amplitudes of scalar perturbation $A_s=2.105\pm 0.030 \times 10^{-9}$ and tensor perturbation $A_t<10^{-10}$, constrained by Planck-2018 \cite{Planck:2018jri}. Applying the normalization formula \cite{Sarkar:2023cpd}, we can obtain the inflationary energy scale $(V_0)^{1/4}=2.36\times 10^{-3}$ in the reduced Planck unit, or, $(V_0)^{1/4}=5.74\times 10^{15}$ GeV, for $\alpha=\frac{1}{3}$ and $N\approx 60$, which lie under the Planck bounds \cite{Planck:2018jri,Planck:2018vyg}. For 
$N=60$, we get from Eqs. (\ref{eq:12}) and (\ref{eq:13}): $n_s=0.967$, $r=1.11\times 10^{-3}$ and $n_t=-\frac{r}{8}=-1.387\times 10^{-4}$, which are consistent with the Planck bounds. However to estimate the mode-dependent values of the cosmological parameters around $k=0.002$ Mpc$^{-1}$ by the method described in Ref. \cite{Sarkar:2021ird}, the form of $V(\mathcal{R})$ of Eq. (\ref{eq:10}) should be exactly known. This amounts to a precise computation of $V_F(\langle\mathcal{V}\rangle,\mathcal{R})$ and $V_D(\langle\mathcal{V}\rangle)$ by the process described in Ref. \cite{Let:2023dtb}. Derivation of $V_F$ requires calculation of the contravariant K\"{a}hler metric $\mathcal{K}^{\alpha\Bar{\beta}}$, which can be expressed in block matrix form as
\begin{equation}\small\small
    \mathcal{K}^{\alpha\Bar{\beta}}=\left(\begin{array}{c|c}
       \mathcal{K}^{\rho_i\Bar{\rho}_j}& \mathcal{K}^{\rho_i\Bar{\Phi}_k} \\\hline
        \mathcal{K}^{\Phi_k\Bar{\rho}_j} & \mathcal{K}^{\Phi_k\Bar{\Phi}_{k'}} 
        \label{eq:matrix}
    \end{array}\right),
\end{equation} where, $i,j=1,2,3$ and $k,k'=1,2,3,\cdot\cdot\cdot N$. Therefore $\mathcal{K}^{\alpha\Bar{\beta}}$ is a $(N+3)\times (N+3)$ matrix. Now as explained earlier, implementation of the quantum corrections, in a controlled way, requires a large value of $N$ ($\sim 10^4$). Therefore direct computation of the K\"{a}hler metric elements \textit{vis-\`{a}-vis} the potential and then minimizing it, is a formidable task. Some publicly available codes \cite{Dias:2015rca,Ronayne:2017qzn,Butchers:2018hds} may mitigate this problem. However, in this paper, our effort will be to make a schematic analysis of the potential which, as we shall show, will be able to extract important information about it.
\par Interestingly, the values of $n_s$ and $r$ found from the generic potential of Eq. (\ref{eq:10}) are almost the same as that from $T$ or $E$ models, for $\alpha=1/3$, at $k=0.002$ Mpc$^{-1}$, obtained by mode analysis in Ref. \cite{Sarkar:2021ird}. Therefore, the values of the cosmological parameters do not depend on any specific form of the potential. This is because of the attractor property of the potential, which implies that irrespective of the shape of the potential, cosmological parameters will uniquely be determined by the order of the leading pole and the coefficient of the term containing the leading pole in the Laurent expansion of the kinetic factor of the Lagrangian (which are $2$ and $\frac{1}{2}$ here, respectively). However the form of the potential is important in knowing the origin of the generic slow-roll plateau, which makes the attractor potentials different from other models. \par Now, as mentioned above, we analyze the potential in this way. We have,
\begin{equation}\small\small
    V\left(\sqrt{2\langle\tau_3\rangle}\tanh{\frac{\phi}{\sqrt{2}}}\right)=V\left(\sqrt{2\langle\tau_3\rangle}\frac{1-e^{-\sqrt{2}\phi}}{1+e^{-\sqrt{2}\phi}}\right).
\end{equation} Choosing $x=e^{-\sqrt{2}\phi}$, we get,
\begin{equation}\small\small 
\begin{split}
    &V\left(\sqrt{2\langle\tau_3\rangle}\tanh{\frac{\phi}{\sqrt{2}}}\right)\\&=V\left(\sqrt{2\langle\tau_3\rangle}\left(\frac{1-x}{1+x}\right)\right)=f(x)\quad\quad (\mathrm{say}).
\end{split}
\end{equation}
The pole at positive direction occurs at $\mathcal{R}=\sqrt{2\langle\tau_3\rangle}$, which corresponds to $\phi=\infty$, following the canonical normalization. Therefore, here, $x=0$ indicates the position of that pole. Now, expanding $f(x)$ by Taylor series about the pole at $x=0$ we get,
\begin{equation}\small\small
   f(x)=f(0)+x\partial_x f(x)|_{x=0}+\frac{x^2}{2!}\partial_{xx}f(x)|_{x=0}+\cdot\cdot\cdot,
   \label{eq:20}
\end{equation}
where the first term $f(0)$ corresponds to $V\left(\sqrt{2\langle\tau_3\rangle}\right)=V_0$, the scale of inflation. If $c_n=\frac{1}{n!f(0)}\frac{\partial^n}{\partial x^n}f(x)|_{x=0}$, then neglecting the terms $\mathcal{O}(x^3)$ and keeping only the dominating terms we obtain
\begin{equation}\small\small
    f(x)\approx V_0\left(1+c_1x+c_2x^2\right).
    \label{eq:Last}
\end{equation}
The inflaton potential can now be obtained if we again transform $x\rightarrow\phi$ by putting $x=e^{-\sqrt{2}\phi}$ in Eq. (\ref{eq:Last}) as,
\begin{equation}\small\small
\begin{split}\small\small  &V\left(\sqrt{2\langle\tau_3\rangle}\tanh{\frac{\phi}{\sqrt{2}}}\right)=\\& \boxed{V_{\mathrm{inf}}(\phi)=V_0\left(1+c_1e^{-\sqrt{2}\phi}+c_2e^{-2\sqrt{2}\phi}\right)},
\end{split}
\label{eq:22}
\end{equation}
which is an $\alpha$-attractor potential with $\alpha=\frac{1}{3}$. Thus we notice that, although a rigorous derivation of the inflaton potential is untenable, we can at least estimate the approximated version shown in Eq. (\ref{eq:22}) through the process described above.\begin{figure*}
       \centering
    \onefigure[width=0.5\linewidth]{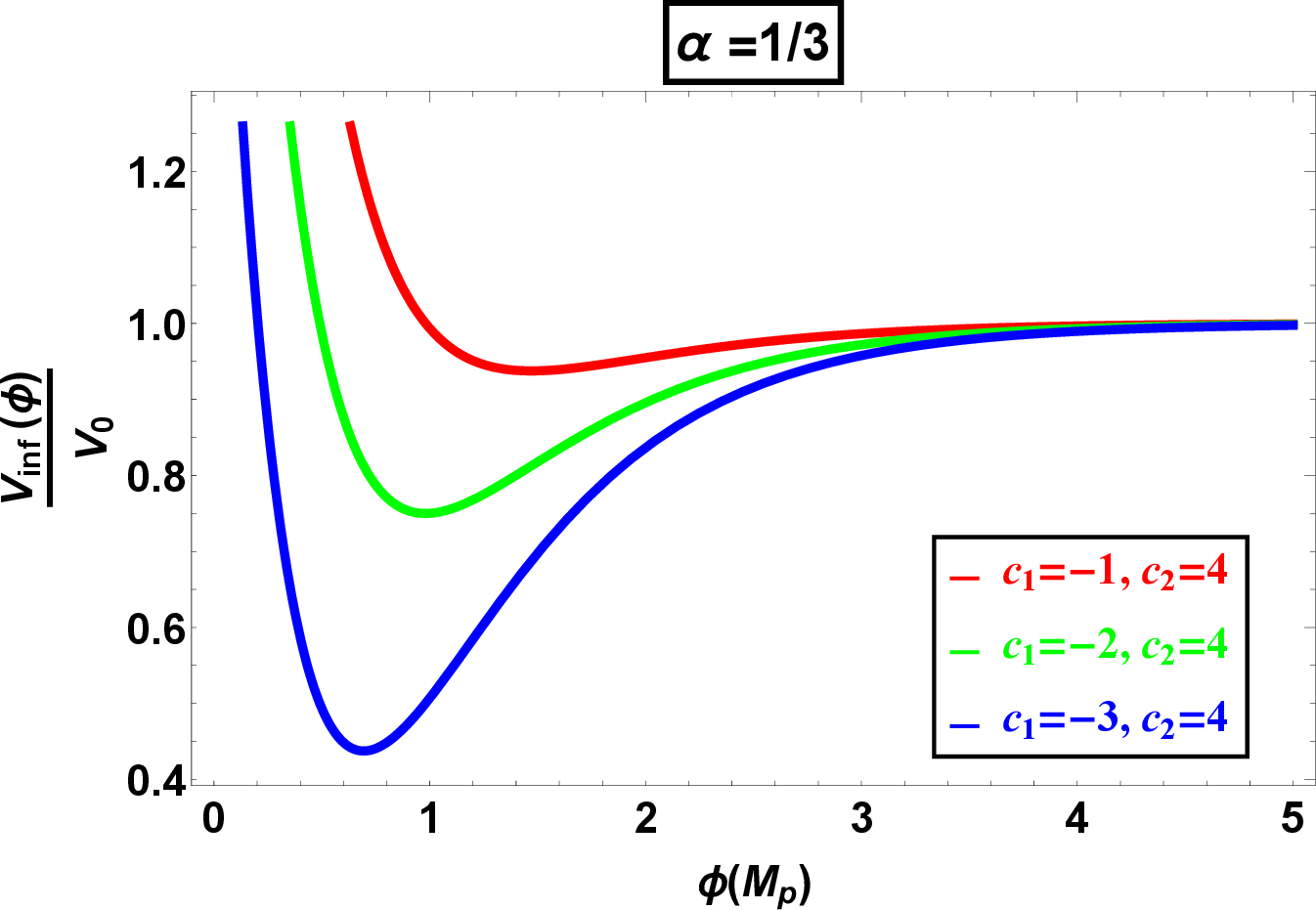}
    \caption{String theoretically derived slow-roll $\alpha$-attractor inflaton potential for three values of $c_1$ and one value of $c_2$.}
    \label{fig:fig1}  
   \end{figure*}
   \begin{figure*}
       \centering
    \onefigure[width=0.5\linewidth]{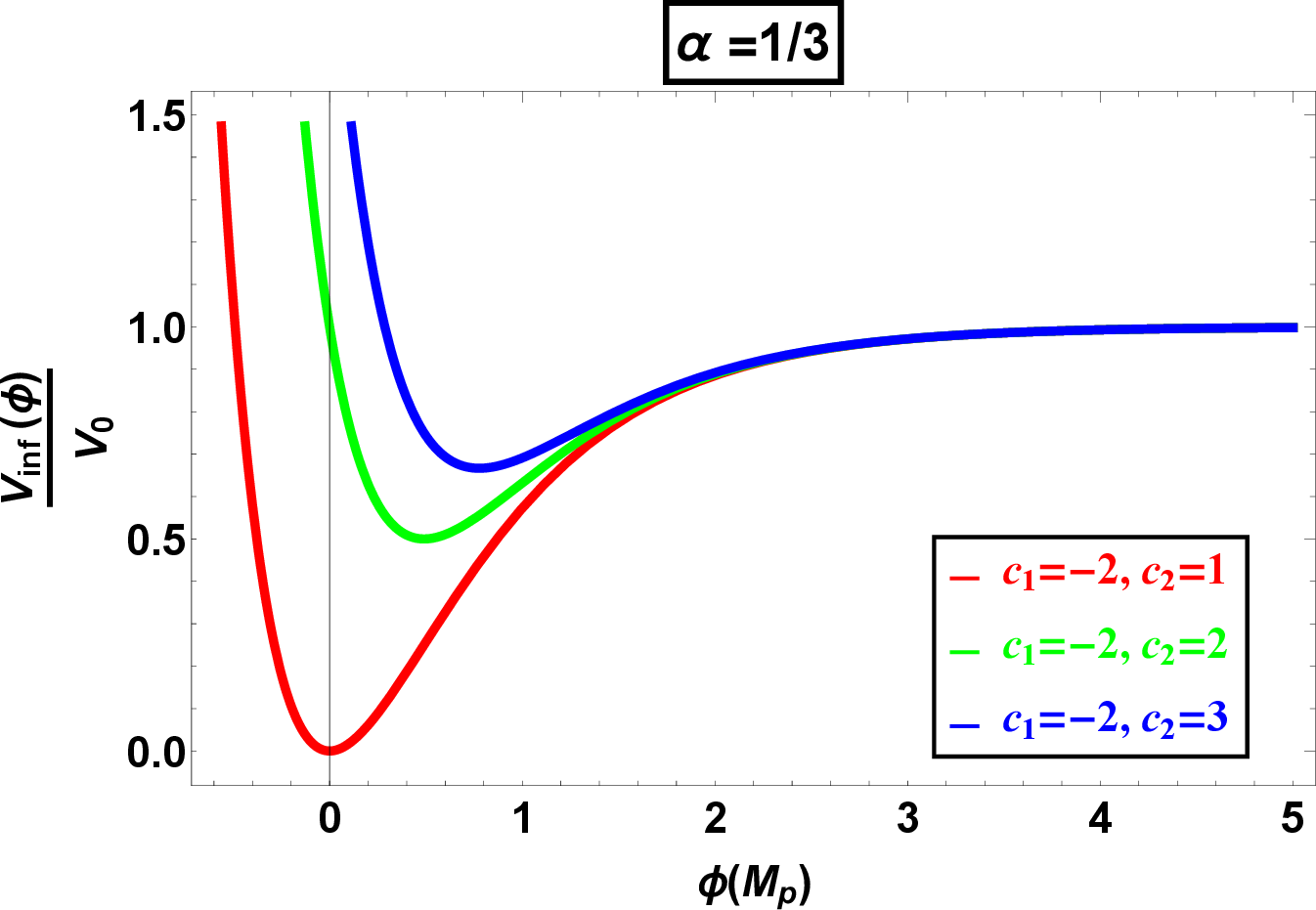}
    \caption{String theoretically derived slow-roll $\alpha$-attractor inflaton potential for three values of $c_2$ and one value of $c_1$.}
    \label{fig:fig2} 
   \end{figure*}
   \begin{figure*}
        \centering
    \onefigure[width=0.5\linewidth]{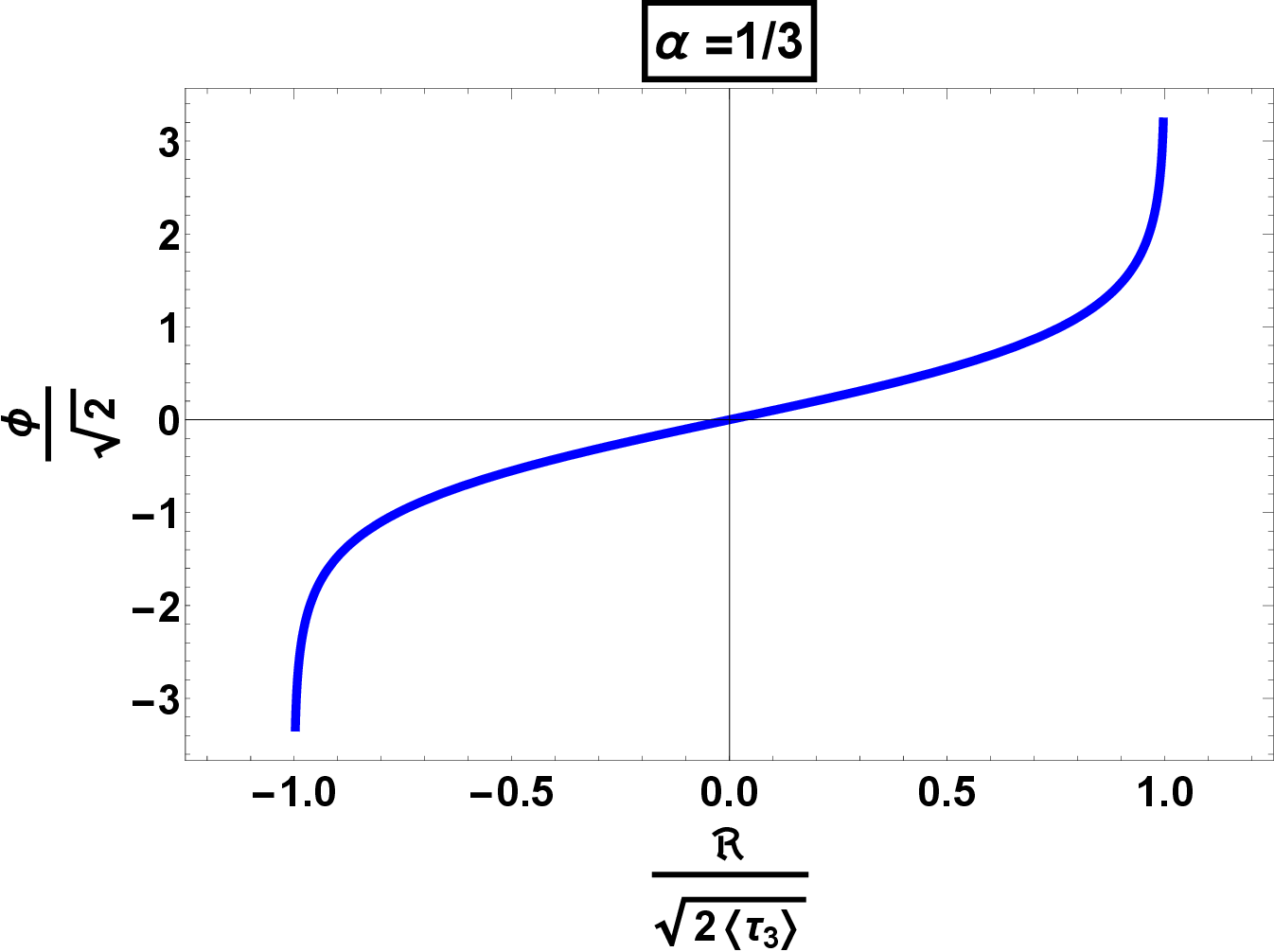}
    \caption{Asymptotic pole behavior of the inflaton field.}
    \label{fig:fig3} 
   \end{figure*}In figures \ref{fig:fig1} and \ref{fig:fig2} the potential of Eq. (\ref{eq:22}) is plotted against $\phi$, measured in reduced Planck mass, for different values of $c_1$ and $c_2$. These constants depend upon various string theoretic parameters related to the moduli stabilizations. Let, at $\phi=\phi_{\mathrm{min}}$, $V_{\mathrm{inf}}(\phi)$ attains the minimum (or the vacuum), \textit{i.e.} $V_{\mathrm{inf}}'(\phi_{\mathrm{min}})=0$ and $V_{\mathrm{inf}}''(\phi_{\mathrm{min}})>0$, where prime denotes derivative with respect to $\phi$. We get, here, \begin{equation}\small
       \phi_{\mathrm{min}}=\frac{1}{\sqrt{2}}\ln{\left(-\frac{2c_2}{c_1}\right)}
       \label{eq:30}
   \end{equation} and 
   \begin{equation}\small
       V_{\mathrm{inf}}''(\phi_{\mathrm{min}})=V_0\frac{c_1^2}{c_2}.
   \end{equation}
   Therefore at $\phi_{\mathrm{min}}$,
   \begin{equation}\small
       V_{\mathrm{inf}}(\phi_{\mathrm{min}})=V_0\left(1-\frac{c_1^2}{4c_2}\right)
       \label{eq:32}
   \end{equation} for $\phi_{\mathrm{min}}>0$ and $V_{\mathrm{inf}}''(\phi_{\mathrm{min}})>0$, which is possible only when $c_1<0$ and $c_2>0$. For example, if $c_1=-2$ and $c_2=1$, then $V_{\mathrm{inf}}(\phi)$ is the $E$ model $\alpha=1/3$ -- attractor. However, in both cases, an uplifting and a shifting (towards the right) of the vacuum are identified for variations of $c_{1,2}$, which can also be understood from Eqs. (\ref{eq:30}) and (\ref{eq:32}).\par
The most crucial observation is that the slow-roll region \textit{i.e.} the infinitely extended flat direction of the canonical inflaton field $\phi$, does not depend upon the fine-tuning of the model parameters $c_1$ and $c_2$. The plateau portion is a generic property of the attractor potential. It arises during field redefinition from non-canonical ($\mathcal{R}$) to canonical ($\phi$) one by $\mathcal{R}=\sqrt{2\langle\tau_3\rangle}\tanh{\frac{\phi}{\sqrt{2}}}$. Figure \ref{fig:fig3} shows that when $\mathcal{R}$ hits the poles at $\pm \sqrt{2\langle\tau_3\rangle}$, then in canonical space they are stretched to infinity, giving rise to a slow-roll region of the inflaton potential. This pole behavior emerges not only for the potential considered here, but for all $\alpha$-attractors irrespective of their origins. This is the so-called `\textit{asymptotic freedom}' of the inflaton field \cite{Kallosh:2016gqp} near the boundary of the moduli space\footnote{see also Ref. \cite{Linde:2016uec} for similar behaviour in the context of multi-field inflation.}. Such generic slow-roll behavior is responsible for the universal predictions of the cosmological parameters \textit{viz.,} scale of inflation ($V_0^{1/4}$), amplitudes of scalar and tensor perturbations ($A_{s,t}$), spectral indices ($n_{s,t}$) and tensor-to-scalar ratio ($r$). The notable observation, here, is the origin of the underlying quadratic pole structure that we have derived from string compactification. Another important point is that, the inflaton field does not arise from $CY_3$ volume, rather a combination of open string moduli manifests as the inflaton field, which makes this prescription novel. The advantage of such a procedure is the inclusion of the quadratic pole in the Lagrangian, which is generally unusual in the realm of moduli stabilization. However, the stabilized volume $\langle\mathcal{V}\rangle\sim\sqrt{\langle\tau_3\rangle}$ has a significant role in determining the position of the pole in non-canonical field space.\par In the present analysis, we have not considered the contributions of $\Theta_k$ and $\gamma_k$ fields in Eq. (\ref{eq:phimoduli}) to make the calculations tractable and the obtained potential to be single field type, which is desirable. We should note, however, that in that case the inflaton potential will be of multi-field type, which is not generally expected \textit{vis-\`{a}-vis} experimental observations.\par In conclusion, we have explored the microscopic origin of the quadratic pole structure of the $\alpha$-attractor in the K\"{a}hler moduli stabilization of type IIB/F theory. The inflaton field arises from the open string sector of the $CY_3$ moduli spectrum in the radial direction and the resulting inflaton potential is a single field $\alpha=1/3$-attractor having a generic slow-roll plateau, which is responsible for conforming to the experimental bounds and insensitive to the model parameters.\par The obtained $\alpha$-attractor belongs to a landscape of many other possibilities depending upon various compactification schemes used for moduli stabilization. For example, in the Fibre Inflation model \cite{Cicoli:2008gp,Kallosh:2017wku} with elliptic $K3$ or $T^4$ fibration $\alpha=\frac{1}{2}$ and $2$ are found. Also, $\mathcal{N}=8$ maximal supergravity \cite{Ferrara:2016fwe,Kallosh:2017ced} predicts some higher fractional values of $\alpha$ \textit{viz.,} $\alpha=\frac{q}{3}$, $q\in \mathbb{Z}^+$. Interestingly in our framework we found $\alpha=\frac{1}{3}$. However, a precise value of $\alpha$ is robust from the cosmological point of view, specifically as a target of future $B$ mode experiments (see Refs. \cite{Kallosh:2019eeu,Kallosh:2017ced, Sarkar:2021ird} for an extensive review). $B$ modes are the crossed polarization states of the CMB photons emitted from the last scattering surface, which left its imprints in the anisotropies of the CMB radiation. These modes might bring forth conclusive evidences in favor of the inflationary scenario and they can be used to precisely estimate the values of the tensor-to-scalar ratio $r$ and the amplitude of the tensor perturbation $A_t$, which are the characteristics of the primordial gravitational waves. As per Planck-2018, $\alpha=1$ (the Starobinsky limit \cite{Kallosh:2014rga}) for $T$ and $E$ models is sufficient to satisfy the observational bounds for inflation. But, so far as the dark energy (DE) observations are concerned, recent studies (see \cite{Sarkar:2023cpd}, for example) on the quintessential extensions of ordinary $\alpha$-attractors provide some indirect clues for a specific range of $\alpha$-values \textit{viz.,} $0.1\leq\alpha\leq 4.3$, where the lower end manifests to be more effective in deciphering the DE properties. The fractional values of $\alpha$ are the benchmarks to explain the late time expansion of the universe in terms of the dynamical quintessential inflaton field \cite{Dimopoulos:2017zvq}, which are again shown to explain the observed baryon asymmetry of the universe in Ref. \cite{Sarkar:2023gwm}. Even more tiny values $\sim 10^{-3}$ of $\alpha$ have been reported in Ref. \cite{Sarkar:2023vpn} to study an early dark energy motivated inflaton potential in the context of resolving the Hubble tension. Now, which of the value of $\alpha$ will ultimately be settled, is a question of further investigations and we find interesting avenues, (see, for example, \cite{Sarkar:2023cpd,Sarkar:2023vpn, Sarkar:2023gwm}) in this direction.

\acknowledgments
The authors acknowledge the University Grants Commission, The Government of India for the CAS-II program in the Department of Physics, The University of Burdwan. AS acknowledges The Government of West Bengal for granting him the Swami Vivekananda fellowship.

\bibliographystyle{eplbib}
\bibliography{biblio}

\begin{thebibliography}{10}
\expandafter\ifx\csname url\endcsname\relax\def\url#1{\texttt{#1}}\fi

\bibitem{Dodelson:2003ft}
\Name{Dodelson S.} \Book{{Modern Cosmology}} (Academic Press, Amsterdam) 2003.

\bibitem{Planck:2018jri}
\Name{Akrami Y. \etal} \REVIEW{Astron. Astrophys.}{641}{2020}{A10}.

\bibitem{Dimopoulos:2022wzo}
\Name{Dimopoulos K.} \Book{{Introduction to Cosmic Inflation and Dark Energy}} (CRC Press) 2022.

\bibitem{Dimopoulos:2017zvq}
\Name{Dimopoulos K. \and Owen C.} \REVIEW{JCAP}{06}{2017}{027}.

\bibitem{Planck:2018vyg}
\Name{Aghanim N. \etal} \REVIEW{Astron. Astrophys.}{641}{2020}{A6} [Erratum: Astron.Astrophys. 652, C4 (2021)].

\bibitem{Scalisi:2018eaz}
\Name{Scalisi M. \and Valenzuela I.} \REVIEW{JHEP}{08}{2019}{160}.

\bibitem{Kallosh:2013yoa}
\Name{Kallosh R., Linde A. \and Roest D.} \REVIEW{JHEP}{11}{2013}{198}.

\bibitem{Roest:2015qya}
\Name{Roest D. \and Scalisi M.} \REVIEW{Phys. Rev. D}{92}{2015}{043525}.

\bibitem{Galante:2014ifa}
\Name{Galante M., Kallosh R., Linde A. \and Roest D.} \REVIEW{Phys. Rev. Lett.}{114}{2015}{141302}.

\bibitem{Kallosh:2014rga}
\Name{Kallosh R., Linde A. \and Roest D.} \REVIEW{JHEP}{08}{2014}{052}.

\bibitem{Kallosh:2016gqp}
\Name{Kallosh R. \and Linde A.} \REVIEW{JCAP}{06}{2016}{047}.

\bibitem{Kallosh:2015zsa}
\Name{Kallosh R. \and Linde A.} \REVIEW{Comptes Rendus Physique}{16}{2015}{914}.

\bibitem{Carrasco:2015uma}
\Name{Carrasco J. J.~M., Kallosh R., Linde A. \and Roest D.} \REVIEW{Phys. Rev. D}{92}{2015}{041301}.

\bibitem{Carrasco:2015rva}
\Name{Carrasco J. J.~M., Kallosh R. \and Linde A.} \REVIEW{Phys. Rev. D}{92}{2015}{063519}.

\bibitem{Canas-Herrera:2021sjs}
\Name{Ca\~nas Herrera G. \and Renzi F.} \REVIEW{Phys. Rev. D}{104}{2021}{103512}.

\bibitem{Kallosh:2019hzo}
\Name{Kallosh R. \and Linde A.} \REVIEW{Phys. Rev. D}{100}{2019}{123523}.

\bibitem{Kallosh:2019eeu}
\Name{Kallosh R. \and Linde A.} \REVIEW{Phys. Lett. B}{798}{2019}{134970}.

\bibitem{Brissenden:2023yko}
\Name{Brissenden L., Dimopoulos K. \and S\'anchez~L\'opez S.} \REVIEW{Astropart. Phys.}{157}{2024}{102925}.

\bibitem{Dimopoulos:2023tcc}
\Name{Dimopoulos K., Brissenden L. \and S\'anchez~L\'opez S.} \REVIEW{PoSC}{ORFU2022}{2023}{247}.

\bibitem{Sarkar:2021ird}
\Name{Sarkar A., Sarkar C. \and Ghosh B.} \REVIEW{JCAP}{11}{2021}{029}.

\bibitem{Sarkar:2023cpd}
\Name{Sarkar A. \and Ghosh B.} \REVIEW{Phys. Dark Univ.}{41}{2023}{101239}.

\bibitem{Sarkar:2023vpn}
\Name{Sarkar A. \and Ghosh B.} \REVIEW{}{}{2023}{}.

\bibitem{Kachru:2003aw}
\Name{Kachru S., Kallosh R., Linde A.~D. \and Trivedi S.~P.} \REVIEW{Phys. Rev. D}{68}{2003}{046005}.

\bibitem{Antoniadis:2020stf}
\Name{Antoniadis I., Lacombe O. \and Leontaris G.~K.} \REVIEW{Eur. Phys. J. C}{80}{2020}{1014}.

\bibitem{Basiouris:2020jgp}
\Name{Basiouris V. \and Leontaris G.~K.} \REVIEW{Phys. Lett. B}{810}{2020}{135809}.

\bibitem{Basiouris:2021sdf}
\Name{Basiouris V. \and Leontaris G.~K.} \REVIEW{Fortsch. Phys.}{70}{2022}{2100181}.

\bibitem{Let:2022fmu}
\Name{Let A., Sarkar A., Sarkar C. \and Ghosh B.} \REVIEW{EPL}{139}{2022}{59002}.

\bibitem{Let:2023dtb}
\Name{Let A., Sarkar A., Sarkar C. \and Ghosh B.} \REVIEW{EPL}{143}{2023}{39001}.

\bibitem{Dias:2018pgj}
\Name{Dias M., Frazer J., Retolaza A., Scalisi M. \and Westphal A.} \REVIEW{JHEP}{02}{2019}{120}.

\bibitem{Giddings:2001yu}
\Name{Giddings S.~B., Kachru S. \and Polchinski J.} \REVIEW{Phys. Rev. D}{66}{2002}{106006}.

\bibitem{Freedman:2012zz}
\Name{Freedman D.~Z. \and Van~Proeyen A.} \Book{{Supergravity}} (Cambridge Univ. Press, Cambridge, UK) 2012.

\bibitem{Becker:2002nn}
\Name{Becker K., Becker M., Haack M. \and Louis J.} \REVIEW{JHEP}{06}{2002}{060}.

\bibitem{Antoniadis:2018hqy}
\Name{Antoniadis I., Chen Y. \and Leontaris G.~K.} \REVIEW{Eur. Phys. J. C}{78}{2018}{766}.

\bibitem{Antoniadis:2019rkh}
\Name{Antoniadis I., Chen Y. \and Leontaris G.~K.} \REVIEW{JHEP}{01}{2020}{149}.

\bibitem{arfken2011mathematical}
\Name{Arfken G.~B., Weber H.~J. \and Harris F.~E.} \Book{Mathematical methods for physicists: a comprehensive guide} (Academic press) 2011.

\bibitem{Kallosh:2015lwa}
\Name{Kallosh R. \and Linde A.} \REVIEW{Phys. Rev. D}{91}{2015}{083528}.

\bibitem{Dias:2015rca}
\Name{Dias M., Frazer J. \and Seery D.} \REVIEW{JCAP}{12}{2015}{030}.

\bibitem{Ronayne:2017qzn}
\Name{Ronayne J.~W. \and Mulryne D.~J.} \REVIEW{JCAP}{01}{2018}{023}.

\bibitem{Butchers:2018hds}
\Name{Butchers S. \and Seery D.} \REVIEW{JCAP}{07}{2018}{031}.

\bibitem{Linde:2016uec}
\Name{Linde A.} \REVIEW{JCAP}{02}{2017}{028}.

\bibitem{Cicoli:2008gp}
\Name{Cicoli M., Burgess C.~P. \and Quevedo F.} \REVIEW{JCAP}{03}{2009}{013}.

\bibitem{Kallosh:2017wku}
\Name{Kallosh R., Linde A., Roest D., Westphal A. \and Yamada Y.} \REVIEW{JHEP}{02}{2018}{117}.

\bibitem{Ferrara:2016fwe}
\Name{Ferrara S. \and Kallosh R.} \REVIEW{Phys. Rev. D}{94}{2016}{126015}.

\bibitem{Kallosh:2017ced}
\Name{Kallosh R., Linde A., Wrase T. \and Yamada Y.} \REVIEW{JHEP}{04}{2017}{144}.

\bibitem{Sarkar:2023gwm}
\Name{Sarkar A. \and Ghosh B.} \REVIEW{}{}{2023}{}.

\end{thebibliography}

\end{document}